\documentclass[preprint]{aastex631}

\usepackage{ulem}
\usepackage{amsmath}

\usepackage{threeparttable}
\usepackage{hyperref} 
\usepackage{graphicx}
\usepackage{subfigure}
\usepackage{lipsum}
\usepackage{float} 
\usepackage{overpic}
\usepackage{wrapfig} 
\usepackage[version=4]{mhchem} 
\usepackage{listings} 
\usepackage{algorithm,algorithmic} 

\newcommand{\gc}[1]{{\color{blue} #1}}
\usepackage{multirow}

\begin{document}

\title{Diagnosing the particle transport mechanism in the pulsar halo via X-ray observations}

\author[0000-0003-2245-4364]{Qi-Zuo Wu}
\affiliation{Department of Astronomy, Nanjing University, 163 Xianlin Avenue,\\
Nanjing 210023, People’s Republic of China}
\affiliation{Key Laboratory of Modern Astronomy and Astrophysics, Nanjing University, Ministry of Education,\\
Nanjing 210023, People’s Republic of China}

\author[0000-0002-0786-7307]{Chao-Ming Li}
\affiliation{Department of Astronomy, Nanjing University, 163 Xianlin Avenue,\\
Nanjing 210023, People’s Republic of China}
\affiliation{Key Laboratory of Modern Astronomy and Astrophysics, Nanjing University, Ministry of Education,\\
Nanjing 210023, People’s Republic of China}

\author[0000-0003-3089-3762]{Xuan-Han Liang}
\affiliation{Department of Astronomy, Nanjing University, 163 Xianlin Avenue,\\
Nanjing 210023, People’s Republic of China}
\affiliation{Key Laboratory of Modern Astronomy and Astrophysics, Nanjing University, Ministry of Education,\\
Nanjing 210023, People’s Republic of China}

\author[0000-0003-0628-5118]{Chong Ge}
\affiliation{Department of Astronomy, Xiamen University, Xiamen, Fujian 361005, China}

\author[0000-0003-1576-0961]{Ruo-Yu Liu}
\affiliation{Department of Astronomy, Nanjing University, 163 Xianlin Avenue,\\
Nanjing 210023, People’s Republic of China}
\affiliation{Key Laboratory of Modern Astronomy and Astrophysics, Nanjing University, Ministry of Education,\\
Nanjing 210023, People’s Republic of China}

\correspondingauthor{Ruo-Yu Liu}
\email{ryliu@nju.edu.cn}



\begin{abstract}

Pulsar halos (also termed `TeV halo’) are a new class of $\gamma$-ray sources in Galaxy, which manifest as extended $\gamma$-ray emission around middle-age pulsars, as discovered around the Geminga pulsar, the Monogem pulsar and PSR~J0622+3749 by HAWC and LHAASO. A consensus has been reached that the TeV emission comes from the inverse Compton scattering of escaping electrons/positrons from the PWN off soft background radiation field, while the particle transport mechanism in the halo is still in dispute. Currently, there are mainly three interpretations, namely, the isotropic, suppressed diffusion model; the isotropic, unsuppressed diffusion model with considering ballistic propagation of newly injected particles; the anisotropic diffusion model. While the predicted gamma-ray surface brightness profiles by all three models can be more or less consistent with the observation, the implication of the three models for cosmic-ray transport mechanisms and the properties of interstellar magnetic field are quite different. In this study, we calculate the anticipated X-ray emission of pulsar halos under the three models. We show that the synchrotron radiation of these escaping electrons can produce a corresponding X-ray halo around the pulsar, and the expected surface brightness profiles are distinct in three models. We suggest that sensitive X-ray detectors of a large field of view (such as eROSITA and Einstein Probe) with a reasonably long exposure time are crucial to understand the formation mechanism of pulsar halos and serve as a probe to the properties of the interstellar turbulence.

\end{abstract}



\section{Introduction} \label{sec:intro}
    Pulsar halos are a new class of non-thermal radiation sources
    detected by the High-Altitude Water Cherenkov Observatory (HAWC) collaboration a few years ago. Observations show very-high-energy (VHE) $\gamma$-ray emissions up to about 50\,TeV around two middle-aged pulsars, Geminga and Monogem \citep{2017Sci...358..911A}. The $\gamma$-ray morphology exhibits quasi-isotropic characteristics with spatial extension to at least 20\,pc from the pulsar, which is much larger than the size of  the pulsar wind nebula (PWN) of $0.2-0.3$\,pc \citep{2003Sci...301.1345C}. This phenomenon is suggested to occur at the late stage of a pulsar system, typically when the age $T_{\rm age}$ of the pulsar exceeds 100\,kyr. At this stage, the confinement of high-energy electron/positron pairs accelerated inside the PWN becomes weak and these energetic particles can easily escape. Due to the pulsar’s proper motion or the faded supernova remnant (SNR), escaped pairs are diffusing in the ambient interstellar medium (ISM) \citep{2020A&A...636A.113G}. These electrons/positrons ($e^{\pm} $) are supposed to emit nonthermal radiation, from radio to $\gamma$-ray up to hundreds of TeV, via synchrotron radiation and inverse Compton scattering (ICS) off the interstellar infrared (IR) radiation background and the cosmic microwave background (CMB). 

 By employing the isotropic diffusion model (hereafter refer to as the ID Model for simplicity), the surface brightness profile (SBP) can be satisfactorily reproduced. However, the diffusion coefficient, $D(E)$, needs to be at least two orders of magnitude smaller than the typical value in ISM \citep{2009PhRvL.103e1101Y}, which is still lack of convincing explanation. Subsequently, the Large High-Altitude Air
Shower Observatory (LHAASO) reported the detection of another TeV pulsar halo candidate surrounding the middle-aged pulsar PSR~J0622+3749 \citep{2021PhRvL.126x1103A}. The fitting to the SBP and energy spectrum also supports the presence of a suppressed diffusion zone around PSR~J0622+3749, similar to the condition observed in Geminga and Monogem. The existence of the suppressed diffusion zone might originate from the self-generated instability \citep[][but see \citealt{2019MNRAS.488.4074F}]{PhysRevD.98.063017, 2022PhRvD.105l3008M, 2022MNRAS.512..233S}, or implies a different particle propagation mechanism.

\cite{PhysRevD.104.123017} proposed a new diffusion model, considering transition of particle transport from ballistic to diffusive regimes (hereafter, B2D Model). Within one mean free path or a scattering timescale $\tau_{c}= 3D(E)/c^{2}$ after injection, $e^{\pm} $ propagate quasi-ballistically rather than diffusely, and hence the density drops more quickly with increasing distance with respect to the prediction of the ID model at small distance. After several mean free path, the propagation proceeds in the diffusive regime. With this model, an acceptable fit to the $\gamma$-ray data around Geminga, Monogem and PSR~J0622+3749 is obtained with a standard diffusion coefficient in ISM. However, the B2D Model usually requires a higher conversion efficiency $\eta$, even above $100\%$ for some sources \citep{2022ApJ...936..183B}, which could be problematic.

In previous two models, the particle diffusion is isotropic, implying very chaotic configuration of magnetic field in the pulsar halo. \cite{2019PhRvL.123v1103L} proposed that the interstellar magnetic field around these middle-aged pulsars may not be so chaotic. Instead, there could be a non-negligible regular magnetic field component. In this case, $e^{\pm} $ tend to diffuse faster along the mean direction of magnetic field than that perpendicular to it, and the diffusion is be anisotropic. Consequently, the spatial distribution of $e^{\pm}$ is like a cylinder with a symmetric axis aligned with the mean magnetic field direction. Numerical calculations show that considering the anisotropic diffusion (hereafter, the AD model) of particles can reproduce the TeV morphology of Geminga  if the mean magnetic field is roughly aligned with the line of sight. In other words, the slow diffusion may be attributed to the perpendicular diffusion. However, its consistency with the general TeV halo observations is still under debate \citep{2022ApJ...935...65Y,2022PhRvD.106l3033D}.

In general, all three models can more or less explain HAWC and LHAASO's measurements, and it is hard to distinguish them only based on TeV $\gamma$-ray observations of relatively poor angular resolutions. However, the particle transport mechanism is crucial to understand the physical nature of pulsar halo. Since $e^{\pm}$ can also emit X-ray emission via synchrotron radiation in the interstellar magnetic field. X-ray instruments, which usually have advanced angular resolutions, may provide crucial information about these halos. In fact, there are already X-ray observations on the inner region of these halos.  \cite{2019ApJ...875..149L} analyzed the data of XMM-Newton and Chandra, and obtain an upper limit for the diffuse X-ray flux in the inner 600'' region around the Geminga pulsar. They found that the magnetic field inside the halo is required to be <1\,$\mu G$ for ID Model. Another study with only $\sim 400-500\,$s exposure of eROSITA on a few pulsar halos (candidates) also failed to capture the corresponding X-ray emission, and put constraints on the strength of interstellar magnetic field. An indication of the existence of X-ray pulsar halo comes from the Chandra's observation on the region around PSR~J1809-1917 \citep{2023ApJ...949...90L}, but due to limited field of view (FoV) of the observation, the pulsar halo origin of the X-ray emission is not concretely confirmed. 

In this study, we revisit interpretations of the extended $\gamma$-ray emissions around Geminga, Monogem, and PSR~J0622+3749 with the ID, B2D, and AD models, and predict the properties of their corresponding X-ray halos. We examine the detectability of these X-ray halos by possible future observations by large-FoV instruments such as eROSITA \citep{2012arXiv1209.3114M} and the follow-up X-ray telescope of Einstein Probe (EP) \citep{2015arXiv150607735Y},  and whether X-ray observations can distinguish different models. The rest of the paper is organized as follows: in Section 2 we present the history of particle injection in the pulsar system, introduce three diffusion models we used and radiation process of these relativistic particle. In Section 3, we show the X-ray data reduction process and the method to calculate the sensitivity of eROSITA and EP. In Section 4, we show our fitting results towards the Geminga, Monogem and PSR~J0622+3749 based on ID, B2D and AD model respectively. We discuss our results and give a conclusion in Section 5.

\section{Particle transport and radiation}
We know that pulsar is highly-magnetized, rotating neutron star, and its rotational energy is given by
$W_{s}= \frac{1}{2} I \Omega^{2}$, where $I$ is the pulsar's rotation momentum and $\Omega$ is the angular velocity. Generally, $\Omega$ evolves as $\dot{\Omega}=-A \Omega^{n}$, where $n$ is the braking index(assumed to be 3). The age of a pulsar system is given by:
\begin{equation}
t_{\rm age}=\tau_{c}~[1-(\frac{P_{0}}{P})^{2}] 
\end{equation}
where $\tau_{c}= P/2\dot{P}$, $P_{0}$ is the initial rotation period of the pulsar. The rotational energy keeps being dissipated, thus the spin-down luminosity is defined by:
\begin{equation}
L_{\rm s,t}(t) \equiv -I\Omega\dot{\Omega}=\frac{L_{\rm s,0}}{(1+t/\tau_{0})^{ 2  }}
\end{equation}
where $ \tau_{0} \equiv P_{0}/2\dot{P_{0}}= \tau_{c}-t_{\rm age} $ is the initial spin-down time-scale. The physical parameters used are listed in Table.~\ref{tab1}. Part of the spin-down energy will be transferred to $e^{\pm}$, and we adopt a power law with exponential cutoff as the injection spectrum:
\begin{equation}
Q_{\rm inj}(E_{\rm e}, t)  = Q_{0}(t)E_{\rm e}^{-p}e^{E_{\rm e}/E_{\rm cut}}   ,\hspace{3ex} E_{\rm e,min} \leq E_{\rm e}  \leq E_{\rm e,max}       
\end{equation}
where $E_{\rm e,min}$ and $E_{\rm e,max}$ are the minimum and maximum energy of injection spectrum respectively, $p$ is the spectral index and is set to be 1.6 in this work, $E_{\rm cut}$ is the cutoff energy.
After injection, $e^{\pm}$ will not only propagate outward in space but also lose energy through synchrotron radiation and ICS. The energy loss rate is given by \cite{2021Sci...373..425L}:
\begin{equation}
\dot{E}_{\mathrm{e}} = -\frac{4}{3}\sigma_{\mathrm{T}}c\left( \frac{E_{\mathrm{e}}}{m_{\mathrm{e}}c^2} \right)^2\left\{ U_{\mathrm{B}}+\sum_i U_{\mathrm{ph},i}\,/\left[ 1+\left( \frac{2.82kT_{\mathrm{i}}E_{\mathrm{e}}}{m_{\mathrm{e}}^2c^4} \right)^{0.6} \right]^\frac{1.9}{0.6} \right\}
\end{equation}
where $\sigma_{\rm T}$ is the Thomson cross-section,$m_{\rm e}$ is the mass of $e^{\pm}$, $U_{\rm B}=B^2/8\pi$ is the energy density of magnetic field and $U_{{\rm ph},i}$ is the $i$th component of energy density of background photon field. We use the CMB, an infrared radiation field ($T = 30\,$K and $U = 4.8\times10^{-13}\, \rm erg~cm^{-3}$) and a star light radiation field ($T = 5000 \rm K$ and $U =  4.8\times10^{-13}$ $\rm erg$ $\rm cm^{-3}$).

{\bf ID model}: The distribution of $e^{\pm}$ evolves with time both in space and energy space according to following equation:
\begin{equation}
\frac{\partial{N}}{\partial{t}}=\frac{1}{r^2}\frac{\partial }{\partial r}\left(r^2 D \frac{\partial N}{\partial r}\right)- \frac{\partial}{\partial{E_{\rm e}}}(\dot{E_{\rm e}}N)+Q  
\label{propagation}
\end{equation}
where ${D}$ is the diffusion coefficient and has the form in a two-zone ID Model: 
\begin{equation}
D(r,E)=\left\{
\begin{aligned}
&D_{0}~\left(\frac{E}{1 \rm GeV}\right)^{1/2}  ,\hspace{3ex}  0 < r < r_{b}\\
&D_{\rm ISM} ,\hspace{3ex} r \geq  r_{b}
\end{aligned}
\right.
\end{equation}
where $D_{0}$ is subject to the SBP fitting, and $D_{\rm ISM}$ is fixed at $10^{28}(E/{\rm 1\,GeV})^{1/2} \rm cm^2/s$ and $r_{b}=50~\rm pc$ in our work. The two-zone diffusion equation could be solved analytically, given by \cite{2020JPhCS1697a2009O}.

{\bf B2D model}: If the diffusion coefficient is large, Eq.~(\ref{propagation}) faces the superluminal propagation problem, which can be solved by adopting the J$\rm\ddot{u}$ntter function to describe the propagation of $e^{\pm}$ \citep{2009ApJ...693.1275A}. In this work, we follow the treatment proposed by  \cite{PhysRevD.104.123017} to calculate particle distribution in this model. More specifically, it assumes that $e^{\pm}$ is in the ballistic regime within $\tau_{c}$ after being injected and then, as time passes, the multiple deflections experienced in the turbulent magnetic field lead to the isotropization of the particle directions. The contribution of $e^{\pm}$ density is divided into $e^{\pm}$ injected recently in ballistic regime and others in diffusive regime, in which case an angular distribution function of $e^{\pm}$ in the transition from the ballistic regime to diffusive regime needs to be considered \citep{2010PhRvD..82d3002A,2015PhRvD..92h3003P}:
\begin{equation}
    M(\mu)=\frac{1}{Z(x)} {\rm exp} \left(- \frac{3(1-\mu)}{x} \right)
\end{equation}
where $Z(x)= \frac{x}{3}(1-{\rm exp}(-6/x)), x(E)=rc/D(E)=3r/\lambda_{c}, \mu = (l\cos( \theta)-s)/r, r(s,\theta)= \sqrt{l^2+s^2-2ls\cos\theta}$ and $s$ being the distance along the line of sight, $l$ being the distance from the source to the earth, $\theta$ being the angle between the source and the line of sight. Thus the total $e^{\pm}$ density is given by:
\begin{equation}
    N( E_{\rm e}, s, \theta) = ( f_{\rm bal} (E_{\rm e}, s, \theta) + f_{\rm dif}(E_{\rm e}, s, \theta) ) M(\mu(s, \theta) / 2 \pi
\end{equation}
where $f_{\rm bal}$ and $f_{\rm diff}$ are the contribution of ballistic regime and diffusive regime respectively. In B2D model, the diffusion coefficient is assume to be in the form of  $D(E)=D_{\rm 0} (\frac{E}{1 \rm GeV})^{1/2}$ ,where $D_{0}$ is the free parameter.

{\bf AD model}: the magnetic field in ISM has a mean direction with a coherent length, which is typically 50-100 pc and close to the size of pulsar halo. Since $e^{\pm}$ diffuse faster along the mean magnetic field rather than perpendicular to the mean magnetic field, the distribution of $e^{\pm}$ will be anisotropic. The vertical component of diffusion coefficient is $ D_{\perp}= D_{\parallel}M_{A}^4$ \citep{2008ApJ...673..942Y}, where $D_{\parallel}$ is the diffusion coefficient parallel to the magnetic field and $M_{A}$ is the Alfv$\rm\Acute{e}$nic Mach number. In the cylinder coordinate with assuming the z-axis to be the direction of the mean magnetic field and particles injected at the location of pulsar, Eq.~(\ref{propagation}) can be rewritten as follows:
\begin{equation}
    \frac{\partial N}{\partial t}= \frac{1}{r}\frac{\partial}{\partial r}(rD_{rr}\frac{\partial N}{\partial r})+ D_{zz} \frac{\partial^2 N}{\partial z^2}- \frac{\partial}{\partial E_{\rm e}}(\dot{E}_{\rm e}N)+ Q(E_{\rm e})S(t)\delta(r)\delta(z)
    \label{propagtion:ani}
\end{equation}
where $D_{zz}=D_{\parallel}=D_{\rm ISM}$ and
$D_{rr}=D_{\perp}=D_{\rm ISM}M_{A}^4$. By applying the operator splitting technique, Eq.~(\ref{propagtion:ani}) can be solved numerically (see \citealt{2019PhRvL.123v1103L} for details).

\begin{table}[htbp]
    \centering
    \begin{threeparttable}
        \caption{The energy loss rate, age, distance to the Earth, and initial spin-down time-scale of Geminga, Monogem, and PSR~J0622 +3749.}
        \label{tab1}
        \begin{tabular}{ccccc}
            \hline
            \hline
            & $\dot{E}$ [erg/s] & $t_{\mathrm{age}}$ [kyr] & $l$ [kpc] & $\tau_{0}$ [kyr] \\
            \hline
            Geminga & $3.25\times10^{34}$ & 342 & 0.25 & 15.2 \\
            Monogem & $3.8\times10^{34}$ & 111 & 0.29 & 11.7 \\
            PSR~J0622 +3749 & $2.7\times10^{34}$ & 208 & 1.6\tnote{*}  & 50.2 \\
            \hline
        \end{tabular}
        \begin{tablenotes}
            \footnotesize
            \item[*] This distance is obtained by the empirical relationship between GeV luminosity and the spin-down luminosity with large uncertainty \citep{2012ApJ...744..105P}.
        \end{tablenotes}
    \end{threeparttable}
\end{table}

After obtaining $e^\pm$ distributions in three models, we follow the analytical formulae given by \citet{2013RAA....13..680F} to simplify the calculation of synchrotron radiation and that given by \citet{2014ApJ...783..100K} for the IC radiation. Finally, the SBP and energy spectrum can be obtained.

\section{X-ray data reduction and sensitivity calculation}
We performed data reduction of the Chandra observation on the PSR~J0622+3749 region and the XMM-Newton observations on the Monogem region following the techniques of \cite{Ge_2016, Ge_2019} to determine the X-ray emission.  Observation information is in Table \ref{tab2}.

\begin{table}[htbp]
    \centering
    \begin{threeparttable}
        \caption{Information of X-ray Observations. }
        \label{tab2}
        \begin{tabular}{cccccc}
            \hline
            \hline
           Instrument &  Obsid & Observation date & Exposure time (s) \\
            \hline
        Chandra   & 12842      &  2011-01-29 & 2028 / 2050    \\
         \hline
        \multirow{3}{*}{XMM-Newton}  &  0762890101 &  2015-09-19 & unused / 130 000 \\
          &  0112200101 &  2001-10-23 & 17000 / 40923   \\
          &  0853000201 &  2019-10-14 & 68000 / 74100   \\
          &  0853000401 &  2019-10-14 & unused / 7100 \\
            \hline
        \end{tabular}
        \begin{tablenotes}
            \footnotesize
            \item[*] The first row is Chandra observation and the others are XMM-Newton observations. The first and the second number in the column of exposure time means the clean exposure time and the total exposure time respectively. Since the clean exposure time of different CCD chips of MOS1 and MOS2 could be a little bit different after {\tt\string mos-filter}, we show the approximate mean value of the clean exposure time. 
        \end{tablenotes}
    \end{threeparttable}
\end{table}

Chandra observation on the PSR~J0622+3749 region was operated in timed exposure Full Frame Very Faint mode, for which the time resolution is 3.24 s. We dealt the observation with Chandra Interactive Analysis of Observations software package (CIAO) version 4.15 as well as Calibration Database (CALDB) version 4.9.8. We used {\tt\string chandra\_repro} to reprocess the data sets to get new level=2 event files. Point sources were detected with {\tt\string wavdetect}\citep{Freeman_2002} and flare background was filtered by {\tt\string deflare}. To subtract the detector quiescent particle background, we reprojected Chandra stowed background file which was rescaled with photons of $9.5-12$\,keV \citep{Hickox_2006}.  Then the clean image was divided by the exposure map and a circular region around pulsar with $r = 6'$ was chosen to calculate the X-ray photon flux. Supposing a power-law photon spectrum with photon index = 2, we can calculate the energy flux in this region. Since the diffuse X-ray emission was not detected due to short exposure time, we  assumed a Poisson error for photons and calculated the $3\sigma$ error of energy flux as the upper limit of diffuse X-ray emission. The photon flux is $\rm 2.91\times10^{-10} \ photons \ cm^{-2} \ s^{-1} \ arcsec^{-2}$, the solid angle of the area is $\rm 4.07\times10^5 \ arcsec^2$ and the maxium exposure map is $\rm 544992 \ cm^2 \ s \ counts \ photon^{-1}$. With photon index($=2$) and absorption\footnote{\url{https://heasarc.gsfc.nasa.gov/cgi-bin/Tools/xraybg/xraybg.pl}} $ N_{\rm H} = 3.14\times10^{21} \ \rm cm^{-2}$, we can get the $3\sigma$ flux upper limit $\rm 1.3\times10^{-13} \ erg \ cm^{-2} \ s^{-1}$.

\begin{figure}[]
    \centering
    \includegraphics[width=0.96\columnwidth]{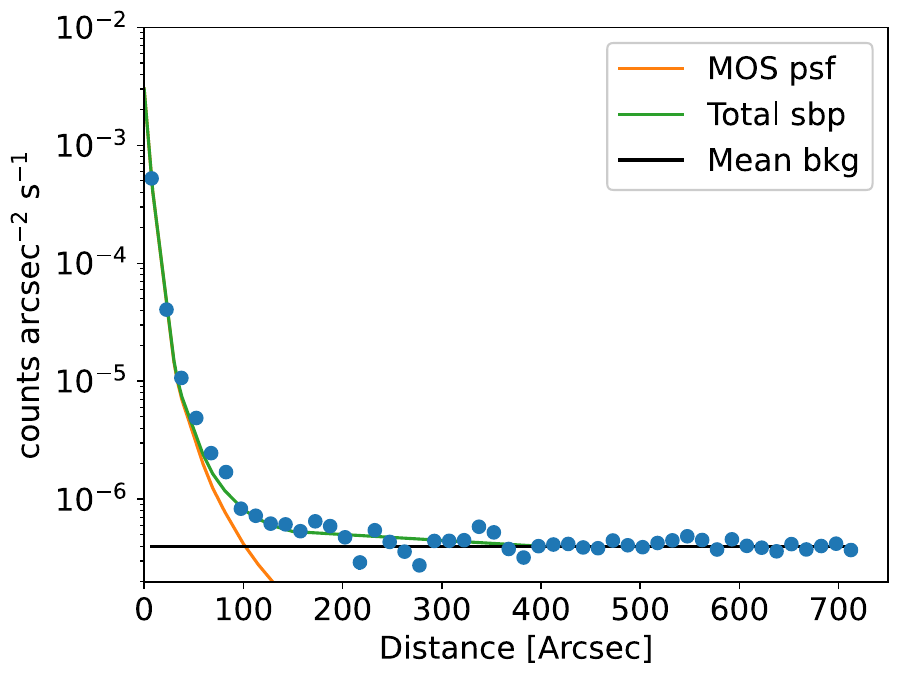}
    \caption{The radial counts profile of the Monogem region centred at the Monogem pulsar observed by XMM-Newton in $0.5-5$\,keV.}
    \label{fig0}
\end{figure}

For XMM-Newton observations, we dealt data using the XMM-Newton Science Analysis System (SAS version 20.0.0) with the associated Current Calibration Files (CCF). MOS data of obsid 0112200101 and 0853000201 were chosen in the downstream processing as other data were either in anomalous state for MOS or small window mode for PN. Following the analysis threads in the XMM-Newton Science Operation center\footnote{\url{https://www.cosmos.esa.int/web/xmm-newton/sas-threads}}, we used {\tt\string mos-filter} to filter out contamination flares of solar soft protons\citep{Read_2003}. Point sources were detected by {\tt\string cheese}. Event images and exposure maps were produced by {\tt\string mos-spectra} while quiescent particle background image was created by {\tt\string mos-back}. We also used {\tt\string proton} to create the residual soft proton contamination image. Then we merged the event and background files as well as exposure maps with {\tt\string merge\_xmm\_comp} and produced a flux image to calculate surface brightness profile in $0.5-5$\,keV (see Figure.~\ref{fig0}). Count rate of the background is $\rm 3\times 10^{-7} \ counts \ s^{-1} \ arcsec^{-2}$, solid angle of the area is $\rm 1.26\times10^6 \ arcsec^2$, and the total clean exposure time(MOS1 + MOS2) is $\rm 170 \ ks$. Using the same method as discussed before, we calculated the $3\sigma$ upper limit of $\rm 3.3\times10^{-14} \ erg \ cm^{-2} \ s^{-1}$.

In this work, we also calculate the sensitivity of eROSITA and EP. We adopt a detection limit of $P_{\rm null} =2.87\times 10^{-7} $, i.e. 5$\sigma$ Gaussian upper-tail probability. For eROSITA, we adopt the background count rate $n_{\rm B}$ and energy conversion factor (ECF) used in \cite{2022MNRAS.513.2884L}: $n_{\rm B}=2.37\times 10^{-3} \ \rm cts \ s^{-1}\ arcmin^{-2}$, ECF=$6.37 \times 10^{11} \ \rm cm^2 \ erg^{-1}$ for Geminga and  $n_{\rm B}=2.45\times 10^{-3} \ \rm cts \ s^{-1}\ arcmin^{-2}$, ECF=$5.91 \times 10^{11} \ \rm cm^2 \ erg^{-1}$ 
 for Monogem. For PSR~J0622+3749, we assume $n_{\rm B}=2\times 10^{-3} \ \rm cts \ s^{-1}\ arcmin^{-2}$ and ECF=$4 \times 10^{11} \ \rm cm^2 \ erg^{-1}$. We calculate the sensitivity of EP following their official Exposure Time Estimator\footnote{\url{http://epfxt.ihep.ac.cn/simulation}}, where the background is taken from \cite{2022APh...13702668Z}. From the website we can derive the exposure time required for $5\sigma$ signal-to-noise ratio for point source, then convert the required time based on the corresponding solid angle. Notice that the website gives the results for one of the two FXT modules, so the predicted exposure time is divided by 
 $\sqrt{2}$. Recently, \citet{2023arXiv231010454K} analyzed eROSITA observations on a few pulsar halos (candidates) with $400-500\,$s for each object, and found no hints of X-ray halos around Geminga and Monogem. We also take their X-ray upper limit into account.

\section{Result} 
We fit SBPs and energy spectra of Geminga's halo and Monogem's halo measured by HAWC and those of PSR~J0622+3749's halo measured by LHAASO with the ID model, B2D model, and AD model respectively. Meanwhile, we use the X-ray upper limits obtained in Section~3 to constrain magnetic field, by requiring the synchrotron radiation of $e^\pm$ not to overshoot the upper limits. The main parameters in the fitting are the magnetic field $B$, the cut-off energy in the $e^\pm$ injection spectrum $E_{\rm cut}$, and the pair conversion efficiency $\eta$. Additionally, the diffusion coefficient has different form in different models. In both ID model and B2D model, $D_{0}$ is treated as a free parameter, noting that $D_0$ in these two models has different definitions. In AD model, the diffusion coefficient parallel to the magnetic field $D_{||}=D_{\rm ISM}$ and the free parameter is the Alfv$\rm\Acute{e}$nic Mach number $M_A$ and the inclination angle between the mean magnetic field and the observer's LOS to the pulsar. We evaluate the goodness of fitting with the $\chi^2$ value based on the SBP fitting.

\subsection{Monogem}
\begin{figure}[]
    \centering
    \subfigure{
    \includegraphics[width=0.32\columnwidth]{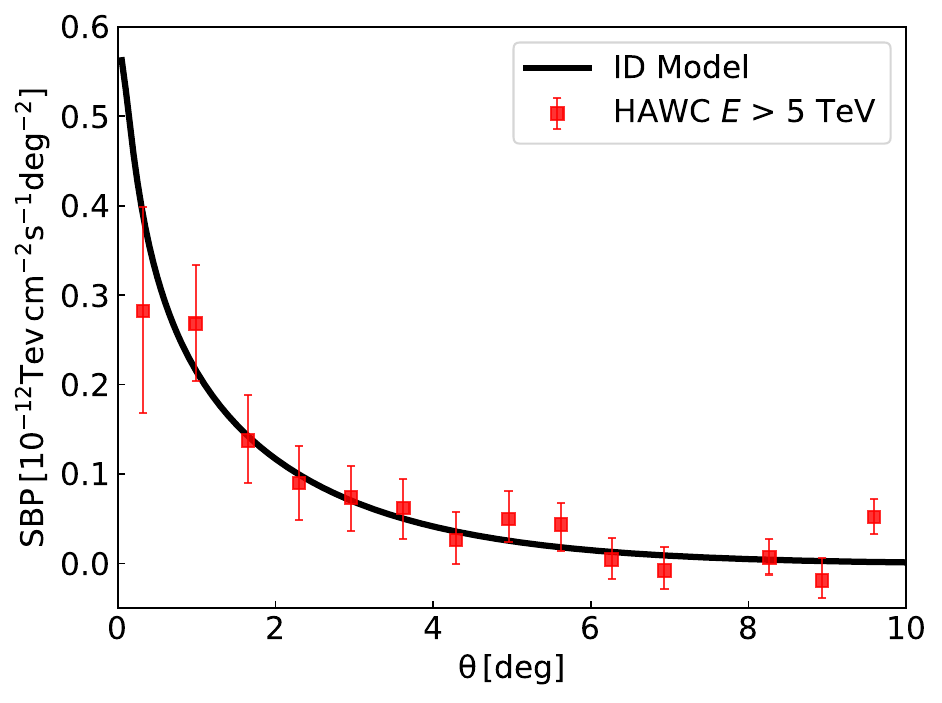}
    \includegraphics[width=0.32\columnwidth]{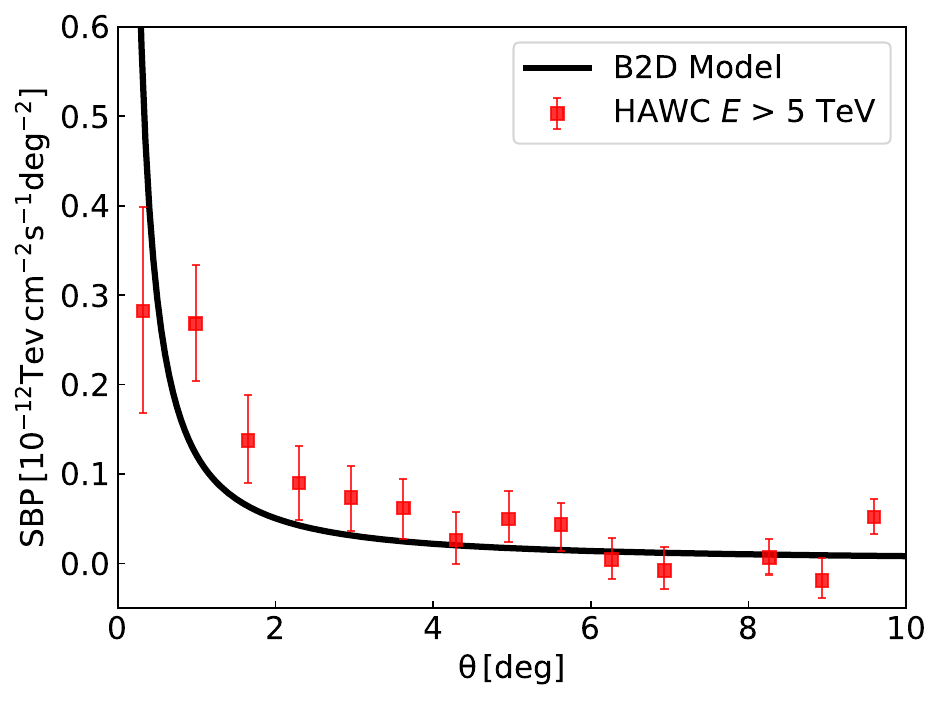}
    \includegraphics[width=0.32\columnwidth]{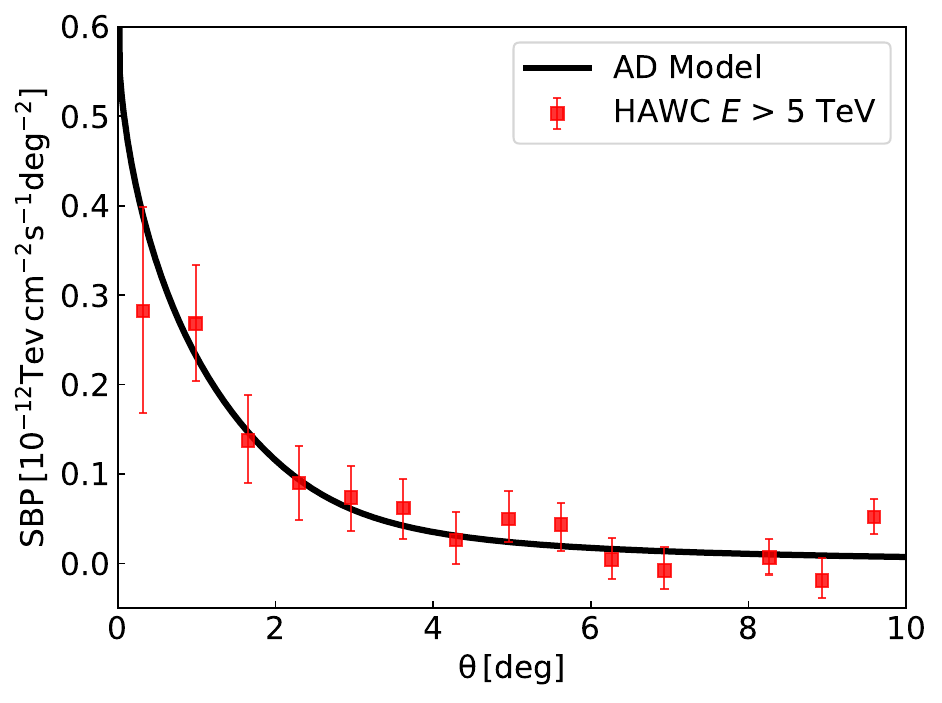}
    }
    \subfigure{
    \includegraphics[width=0.32\columnwidth]{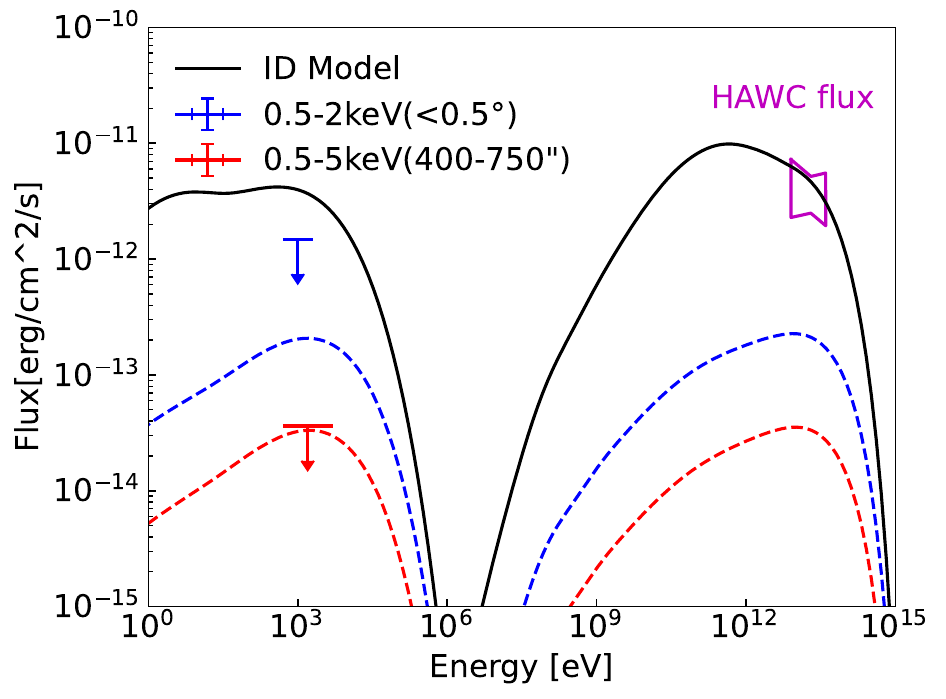}
    \includegraphics[width=0.32\columnwidth]{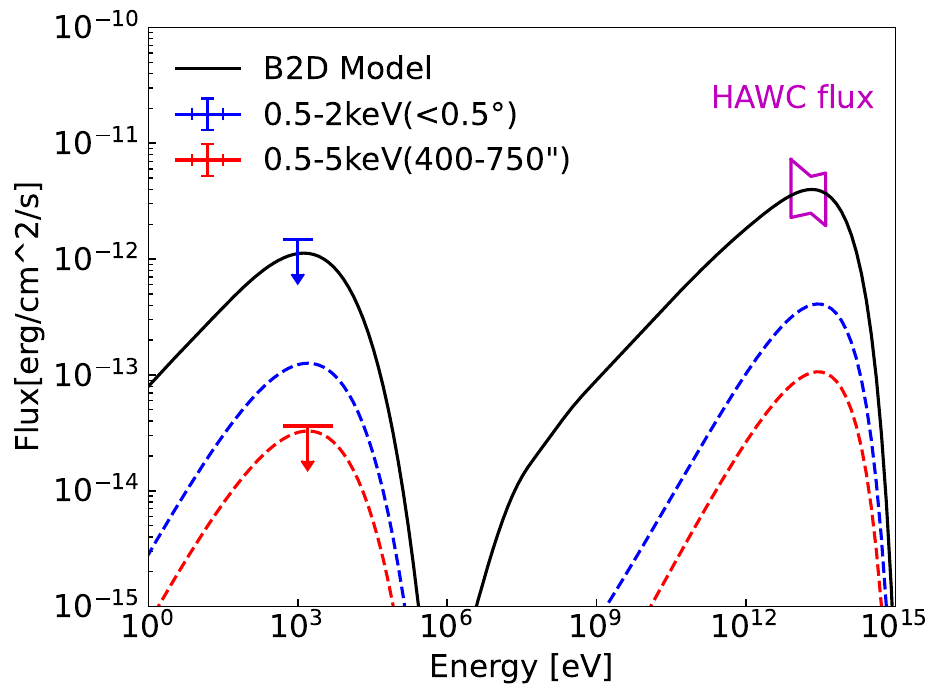}
    \includegraphics[width=0.32\columnwidth]{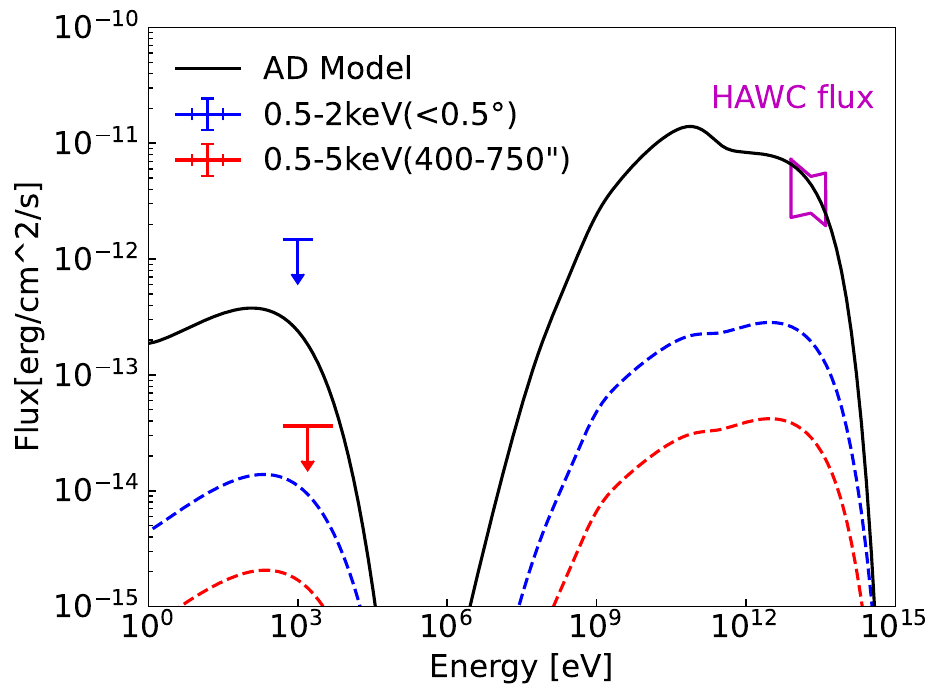}
    }
    \caption{The expected surface brightness profile in TeV and the Energy spectrum of Monogem with ID Model, B2D Model and AD Model. Note that the red and blue dashed lines correspond to the flux of 400--750$^{\prime\prime}$ and within $0.5^{\circ}$ region, respectively.}
    \label{fig1}
\end{figure}

Figure.~\ref{fig1} shows the SBP and energy spectrum of Monogem and Table.~\ref{tab3} shows the fitting parameters. The model is required to fit the HAWC data and satisfy the X-ray upper limit derived at Section 3. For reference, we also show the X-ray upper limit obtained by eROSITA \citep{2023arXiv231010454K} within $0.5^{\circ}$ from the pulsar and the corresponding theoretical flux in the same region. As we can see, the constraint from eROSITA's upper limit is weaker compared to that from XMM-Newton because the exposure of eROSITA is about only 400\,s. For the ID model, the diffusion coefficient needs to be smaller than the typical value in ISM, which is consistent with previous works. The ID model also demands a smaller magnetic  field strength ($B = 2\,\mu$G) to suppress the synchrotron emission. The B2D model gives a steeper TeV profile and a relatively poor fit $\chi^2 = 25.8$. It also demands a smaller magnetic field ($B < 1\,\mu$G) and a high pair conversion efficiency $\eta = 0.76$. The AD model needs the mean magnetic field direction well aligned with the LOS, with the angle between the mean magnetic field and LOS being less than $\phi=10 ^\circ$. Actually, the TeV SBP of Monogem's halo is flatter than that of Geminga's, so a small but none-zero $\phi$ is helpful to flatten the predicted SBP. We see that the ID and AD model both provide a good fit to the HAWC data with similar $\chi^2$ values, while the B2D model gives a poor fit to the TeV SBP. Based on the chi-squared test, the B2D model is less favored but it is hard to rule out it considering the systematic uncertainty of the measurement, which may influence the flux level by 50\% \citep{2017Sci...358..911A}. 

\begin{figure}[]
    \centering
    \includegraphics[width=0.96\columnwidth]{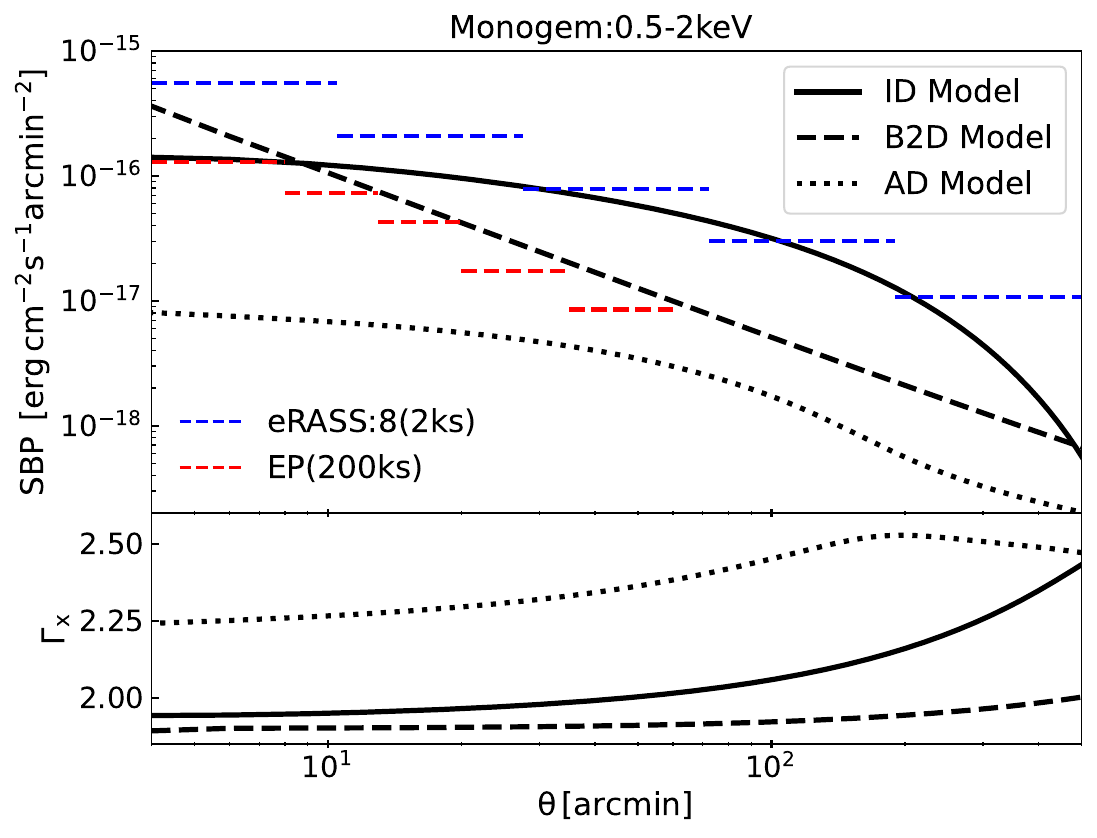}
    \caption{The predicted X-ray intensity and photon index for Monogem with different models. The blue line: the sensitivity of eROSITA all sky survey. The red line: the sensitivity of EP with 200\,ks exposure.}
    \label{fig2}
\end{figure}

Therefore, we calculate the corresponding X-ray SBP of the halo. We exhibit in Figure.~\ref{fig2} the predicted X-ray intensity profile of these models with the same parameters shown in Table.~\ref{tab3}. Differences among the three models are noticeable. The ID model shows obviously upper convex in the shape of the X-ray SBP and the B2D model shows a little lower convex, whereas the AD Model shows upper convex but has a low X-ray intensity. As the dashed blue line shows, the sensitivity of eRASS will be insufficient to distinguish these models. But with a deep exposure (e.g., 200\,ks) towards the central region in the pointing mode, the sensitivity of the EP (the dashed red line) would be enough to diagnose the particle transport model in the Monogem's pulsar halo. We do not look into the expectation of eROSITA with its pointing mode here, but just note that the total eROSITA effective area is about 3.5 times that of EP, so it could reach the same performance with a much shorter exposure time. Our estimate here is optimistic because we use the limiting value of the constrained magnetic field to calculate the synchrotron radiation.

\begin{deluxetable}{ccccccc}
\tablecaption{The diffusion coefficient, magnitude of the magnetic field, conversion eﬀiciency, and cutoff energy of different models for Monogem, PSR~J0622+3749, and Geminga.\label{tab3}}
\tablehead{ \colhead{Source} &\colhead{Model} & \colhead{$D(\rm 1GeV)$ [$\rm cm^2/s$]} & \colhead{$B [\mu G]$} & \colhead{$\eta$} & \colhead{Cutoff energy [TeV]} & \colhead{$\chi^2$}}
\startdata
 Monogem & ID & $1 \times 10^{25}$ & 2 & 0.017 & 150 & 11.7 \\
        & B2D & $3 \times 10^{28}$ & 1 & 0.76 & 100 & 25.8 \\
        &AD & $1 \times 10^{28}, M_{A} = 0.19$ & 5 & 0.033 & 150 & 11.1 \\
\hline 
 PSR~J0622+3749& ID & $2 \times 10^{25}$ & 2.5 & 0.37 & 150 & 9.2 \\
 &B2D& $3 \times 10^{28}$ & 2 & 12.97 & 70 & 18.5 \\
 &AD & $1 \times 10^{28}, M_{A} = 0.2$ & 3 & 0.17 & 250 & 9.5 \\
\hline 
Geminga & ID& $1 \times 10^{25}$ & 0.8 & 0.04 & 100 & 7.0 \\
& B2D& $4 \times 10^{28}$ & 0.4 & 3.04 & 50 & 27.7 \\
&AD & $1 \times 10^{28}, M_{A} = 0.19$ & 5 & 0.04 & 200 & 5.0 \\
\enddata
\end{deluxetable}

\subsection{PSR~J0622+3749}
The angular size of halo of PSR~J0622+3749 is $\lesssim 1^\circ$, which is not much larger than the angular resolution of LHAASO. Therefore, the predicted intensity profile need be convolved with LHAASO's PSF, which can be approximated as a symmetric 2D Gaussian function \citep{2022ApJ...935...65Y}, to compare with the measurement, i.e.,
\begin{equation}
    I( \theta, \varphi)= \  \iint_\Omega I(\theta^{'},\phi^{'}) {\rm PSF}( \theta^{'},\phi^{'}) \, \sin\theta^{'}d\theta^{'}d\phi^{'}
\end{equation}
and then averaged over the azimuthal angle $\phi$ to get the 1D SBP in order to compare with the measured SBP. Note that \cite{PhysRevD.104.123017} convolved the 1D intrinsic theoretical SBP with the 1D Gaussian function of LHAASO's PSF, which is an inaccurate treatment and resulted in a steeper SBP after the convolution than it should be.

\begin{figure}[]
    \centering
    \subfigure{
    \includegraphics[width=0.32\columnwidth]{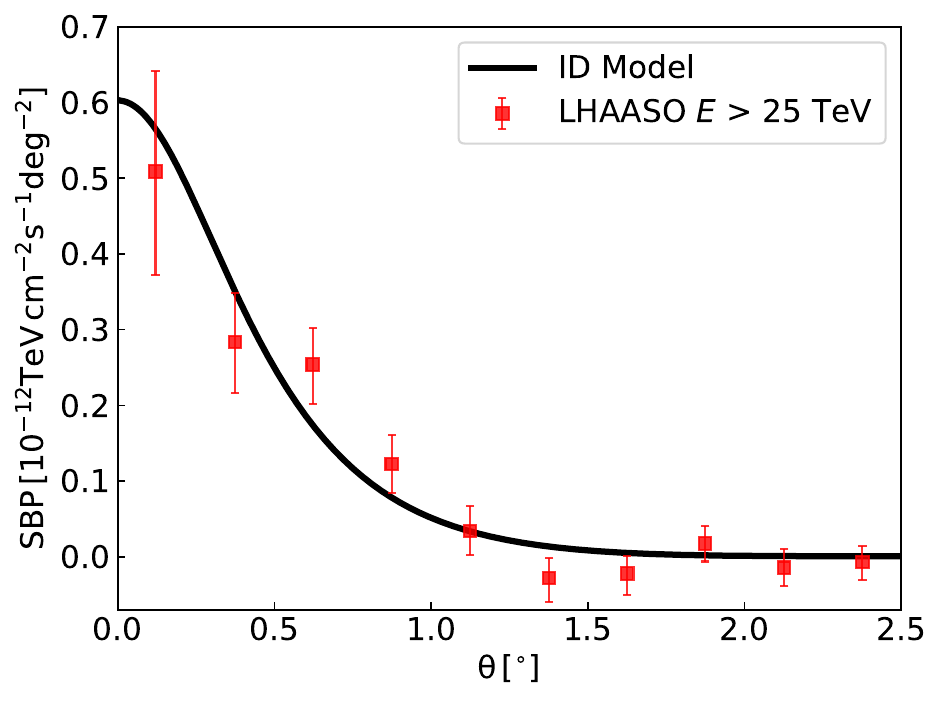}
    \includegraphics[width=0.32\columnwidth]{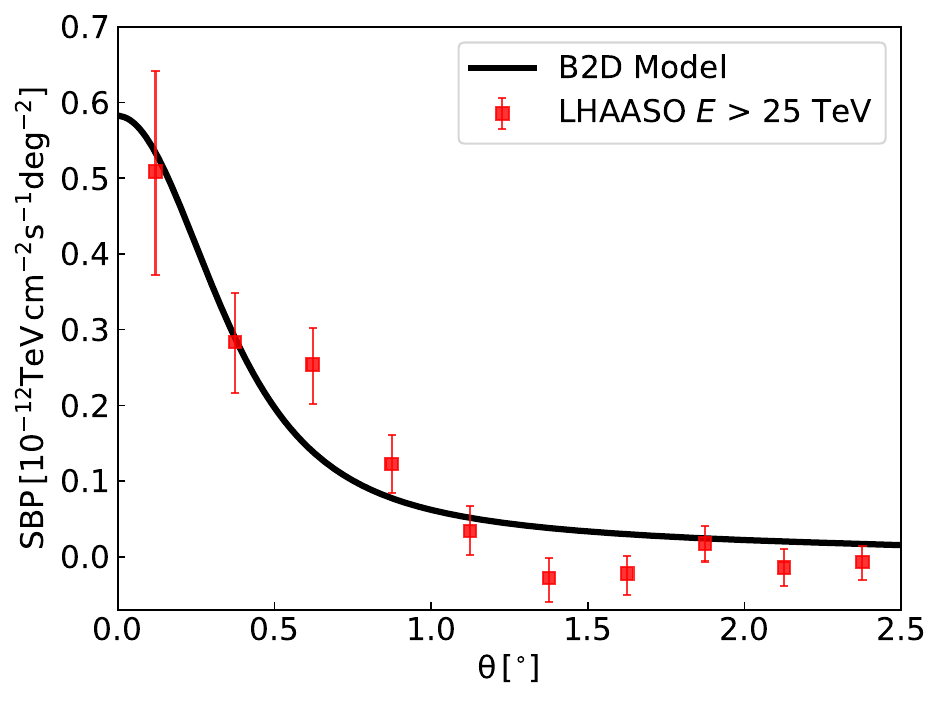}
    \includegraphics[width=0.32\columnwidth]{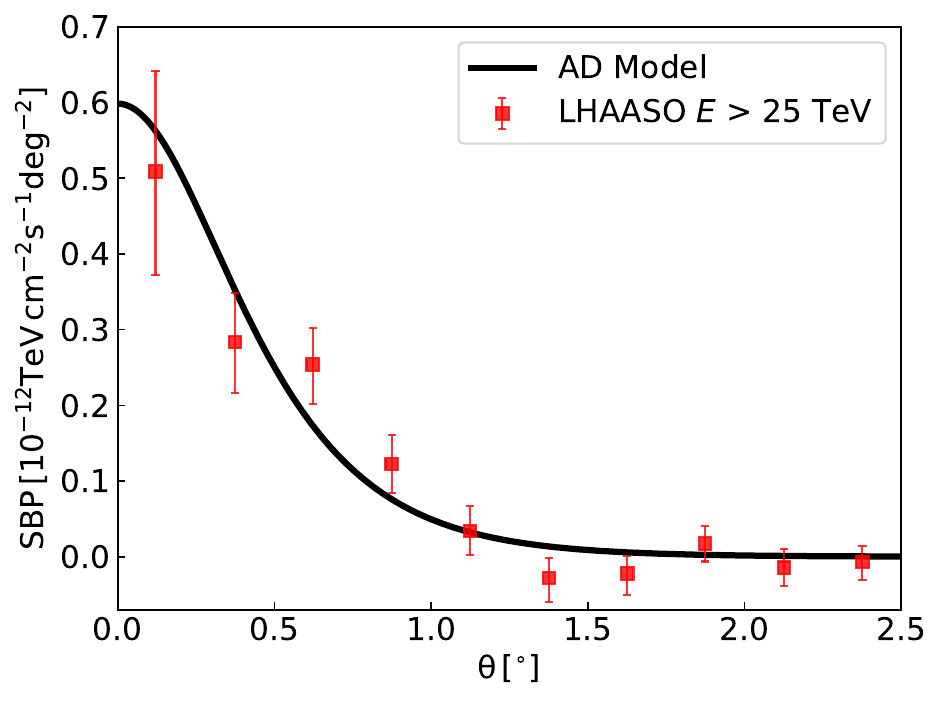}
    }
    \subfigure{
    \includegraphics[width=0.32\columnwidth]{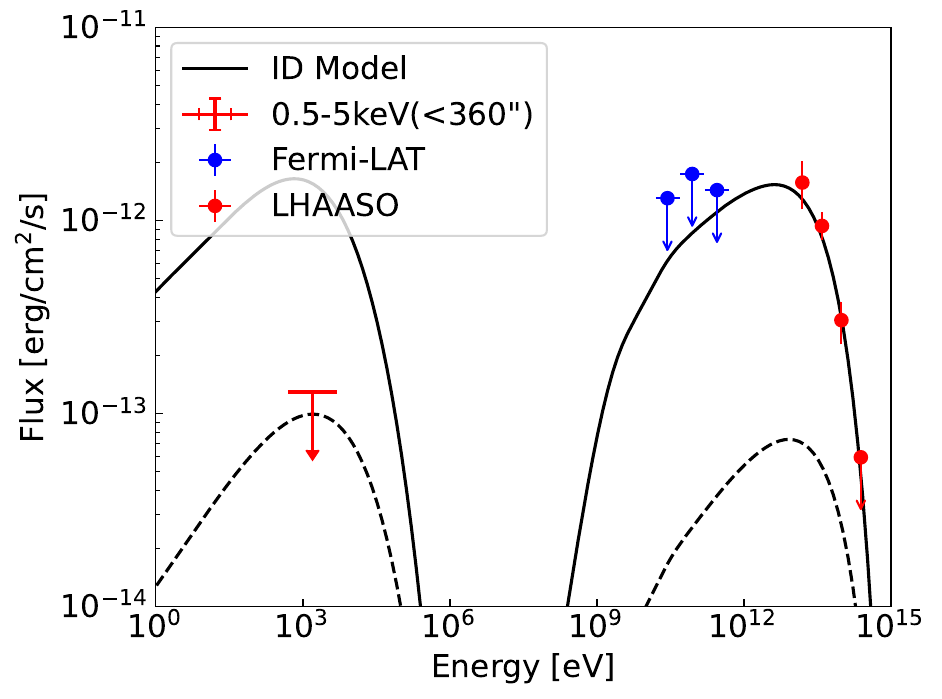}
    \includegraphics[width=0.32\columnwidth]{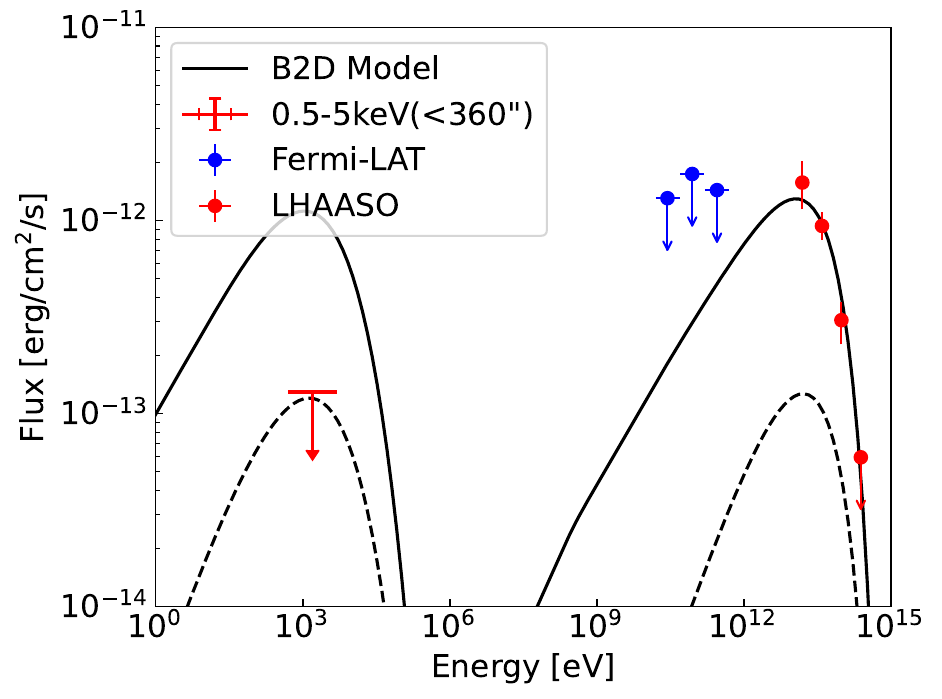}
    \includegraphics[width=0.32\columnwidth]{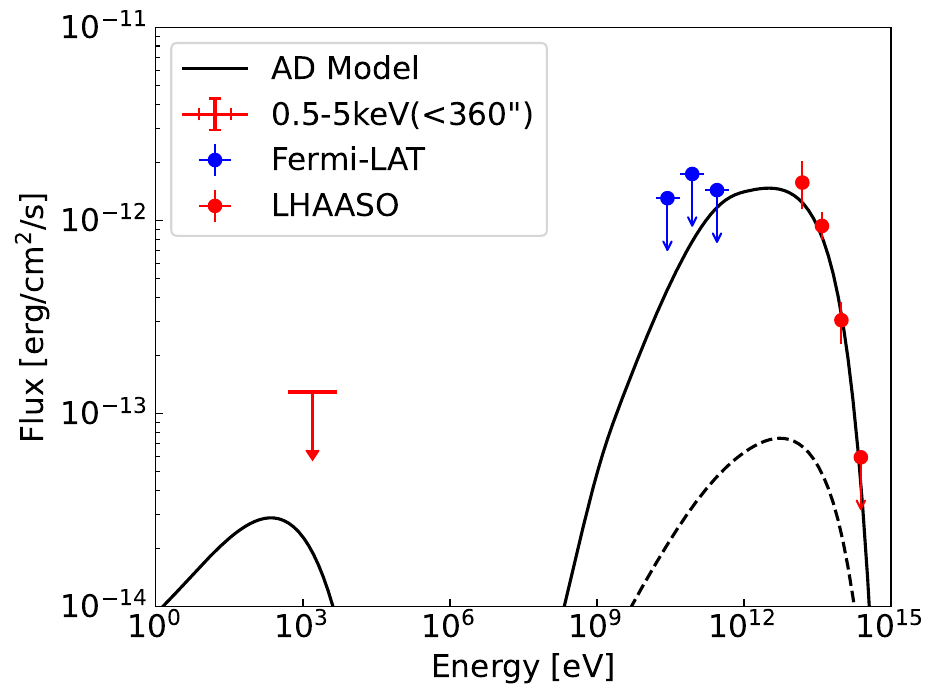}
    }
    \caption{The expected Surface brightness profile in TeV and the Energy spectrum of PSR~J0622+3749 with ID Model, B2D Model and AD Model.}
    \label{fig3}
\end{figure}

After the convolution, the SBPs in TeV are well reproduced by except for B2D Model (see Figure.~\ref{fig3}). The Fermi-LAT upper limit in GeV can be satisfied unless using a very small $P_{0}$, which would lead to too much injection of pairs at early time which remain uncooled today.
Due to the limited exposure time of Chandra towards the PSR~J0622+3749 region, the constraint on the interstellar magnitude field is not so strict (see Table.~\ref{tab3}): $B<2.5\,\mu$G for the ID model and $<2\,\mu$G for B2D Model respectively. Note that, considering the large error bar of the innermost data point in the SBP, the obtained upper limit of the magnetic field may be relaxed by a few times, but the value is not supposed to be much higher than the typical interstellar value i.e., several micro Gauss. 
The diffusion coefficient needed for different models is similar to the Monogem case: a small diffusion zone for ID Model and unsuppressed diffusion coefficient for B2D and AD Model. In AD Model, the included angle $\phi$ is set to be zero given that there is no significant asymmetry in LHAASO observation.

Similar to the case in Monogem, the corresponding X-ray SBPs among three models are distinct. The ID model and the B2D model has  similar level of X-ray intensity but different shape, while the AD model predicts a low X-ray intensity. In such case, a targeted observation with 20\,ks exposure by EP could be enough to distinguish these models. The exposure time is much shorter than that in the Monogem case. This is because the obtained constraint on the magnetic field for PSR~J0622+3749 region is based on only 2\,ks observation of Chandra. And, again, we use the upper limit of the constrained magnetic field to predict the future observation, so the estimation is optimistic. The true magnetic field could be weaker, and the required exposure time to distinguish the ID and B2D models would be correspondingly longer.

\subsection{Geminga}
For Geminga, we employ the X-ray flux upper limit within $600^{\prime\prime}$ from the pulsar, as analyzed by \citet{2019ApJ...875..149L}.
Similar to the previous research, the constraint on the magnetic field in the inner halo is very strict given the several Ms exposure time on this region by Chandra and XMM-Newton, resulting in $B<0.8\,\mu$G for the ID model and $B<0.4\,\mu$G for the B2D model (see Table.~\ref{tab3}). Again, the ID model and AD model give comparatively good fits to the data (see Figure.~\ref{fig5}), while the B2D model yields a worse but acceptable fit considering the systematic error. However, the B2D model is disfavored given the exceptionally large conversion efficiency.

Figure.~\ref{fig6} shows the predicted X-ray intensity, the strong constraint on the X-ray flux leads to the similar intensity for all three models and the X-ray emission is only marginally detectable even with a 1.5\,Ms exposure time of EP. The similar shape of the SBPs predicted by the ID model and the AD model (except at large radius) makes it hard to distinguish these two models, whereas the B2D model could still be distinguished if the X-ray emission is well detected due to the different slope of the SBP. 

\section{Discussion and Conclusion} 


In this study, we adopt a hard injection spectrum with $p=1.6$. Varying the value of $p$ does not have a significant impact on the $\gamma$-ray data fitting, unless a very soft spectrum is employed. Firstly, this is because the spectral index is not accurately determined by HAWC's observation. Besides, the systematic uncertainty in the spectral index is 0.2. For a different $p$, one may choose an appropriate $\eta$ to reproduce the TeV observations well, as shown in \citet{2023PhRvD.107l3020S}. LHAASO's spectral measurement on PSR~J0622+3749's halo is more accurate. However, given the relatively narrow band, one may easily adjust $E_{\rm cut}$ to fit the measured spectrum for a different $p$. Further observations with both LHAASO-WCDA and LHAASO-KM2A could measure a broader band spectrum and provide a better constraint on the spectral index. On the other hand, the X-ray emission arises from the same electrons responsible for the TeV emission. Therefore, the expected X-ray flux is also insensitive to the spectral index, as long as the TeV data is reproduced. 

Another uncertainty is the distance of PSR~J0622+3749. The distance employed in previous calculation, i.e., 1.6\,kpc, is based on an empirical relation between the GeV luminosity and spin-down luminosity of pulsars. We also test the case of $d=800\,$pc for PSR~J0622+3749 and find it unimportant to our results except for a smaller conversion efficiency $\eta$ needed.



The particle transport mechanism of pulsar halos is still in dispute. In this work we revisited the interpretations of the isotropic diffusion model (ID model), the diffusion model considering transition from ballistic to diffuse propagation regime (B2D model) and the anisotropic diffusion model (AD model) for the halos of Geminga, Monogem and PSR~J0622+3749. We calculated spectra and surface brightness profiles of their X-ray halos produced by synchrotron radiation of pairs responsible for TeV emission under the three models. We also analyzed the XMM-Newton data of the Monogem halo and the Chandra data of PSR~J0622+3749's halo, and obtained the flux upper limits in the X-ray band for inner regions of these halos. By confronting the theoretical expectation with the present constraints from the archival X-ray data for the central region of these halos, we may derive constraints on the magnetic field in the halo. While the $\gamma$-ray observation is hard to distinguish these models either through the SBP or the energy spectrum, X-ray SBPs predicted by three models are different. We estimate the detectability of the X-ray counterparts of three halos with the future observations of eROSITA and EP. We found that in the most optimistic case, observations of EP could differentiate these models with 200\,ks exposure on Monogem and 20\,ks exposure on PSR~J0622+3749. Additionally, with multiple pointings of exposure different areas along one radial direction, XMM-Newton and Chandra may also help to differentiate the models. However, the X-ray emission of Geminga's halo is hard to detect given the current constraint on the magnetic field therein. 

\section*{Acknowledgements}
This work is supported by National Natural Science Foundation of China under grants No.~12393852, U2031105 and 12333006.


%

\vspace{5mm}





\bibliography{sample631}{}
\bibliographystyle{aasjournal}



\begin{figure}[]
    \centering
    \includegraphics[width=0.96\columnwidth]{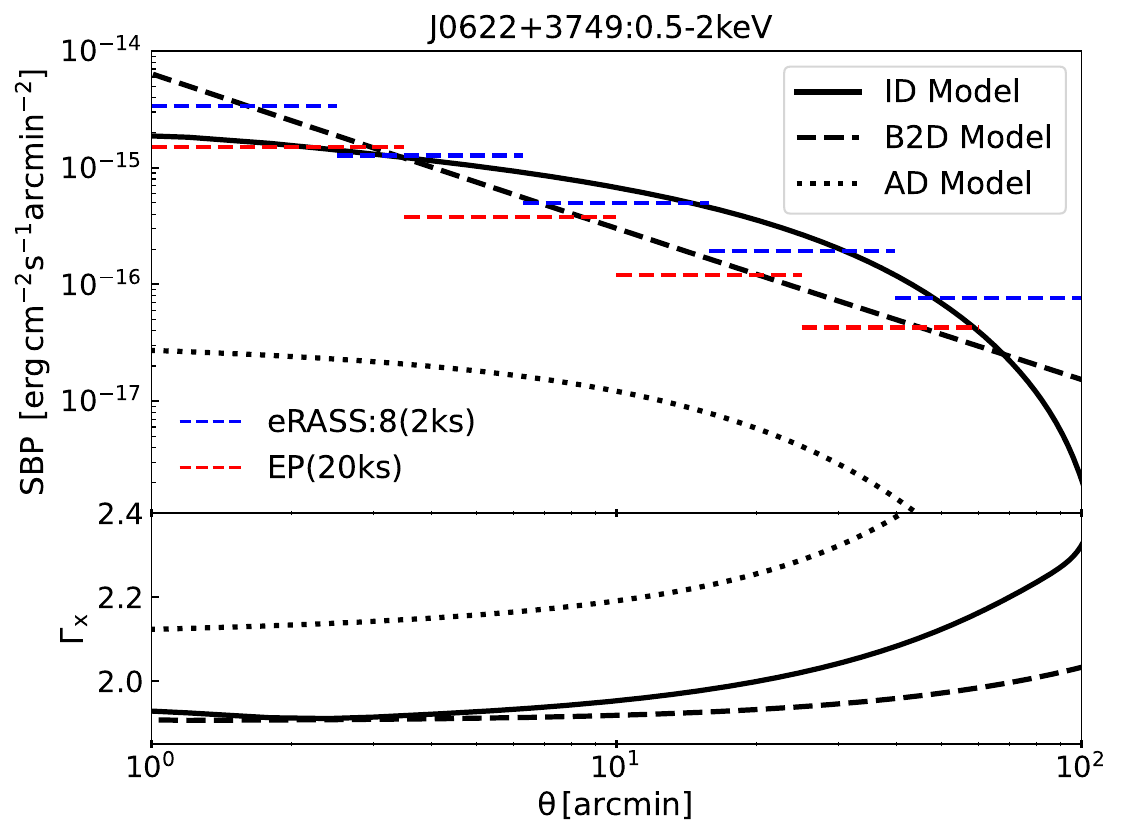}
    \caption{The predicted X-ray intensity and photon index for PSR~J0622+3749 with different models. The blue line: the sensitivity of eROSITA all sky survey. The red line: the sensitivity of EP with 20ks exposure.}
    \label{fig4}
\end{figure}

\begin{figure}[]
    \centering
    \subfigure{
    \includegraphics[width=0.32\columnwidth]{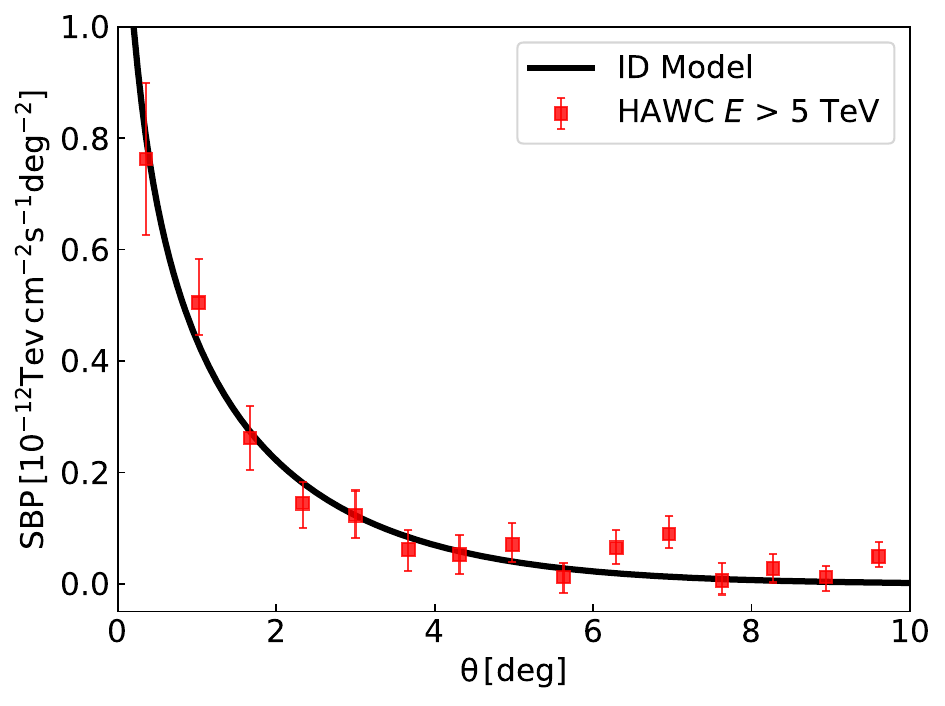}
    \includegraphics[width=0.32\columnwidth]{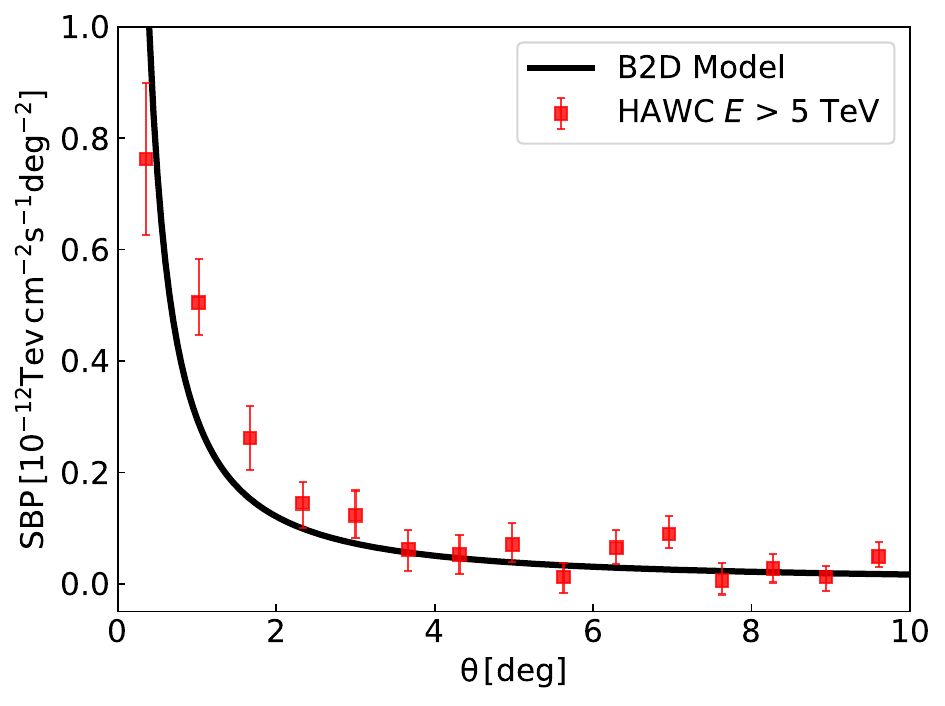}
    \includegraphics[width=0.32\columnwidth]{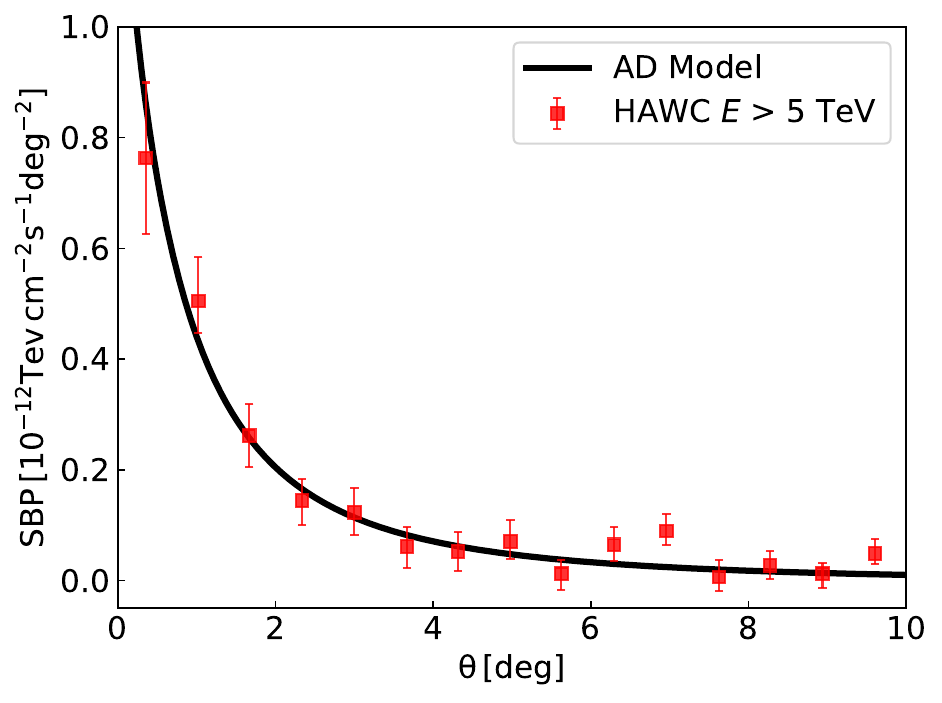}
    }
    \subfigure{
    \includegraphics[width=0.32\columnwidth]{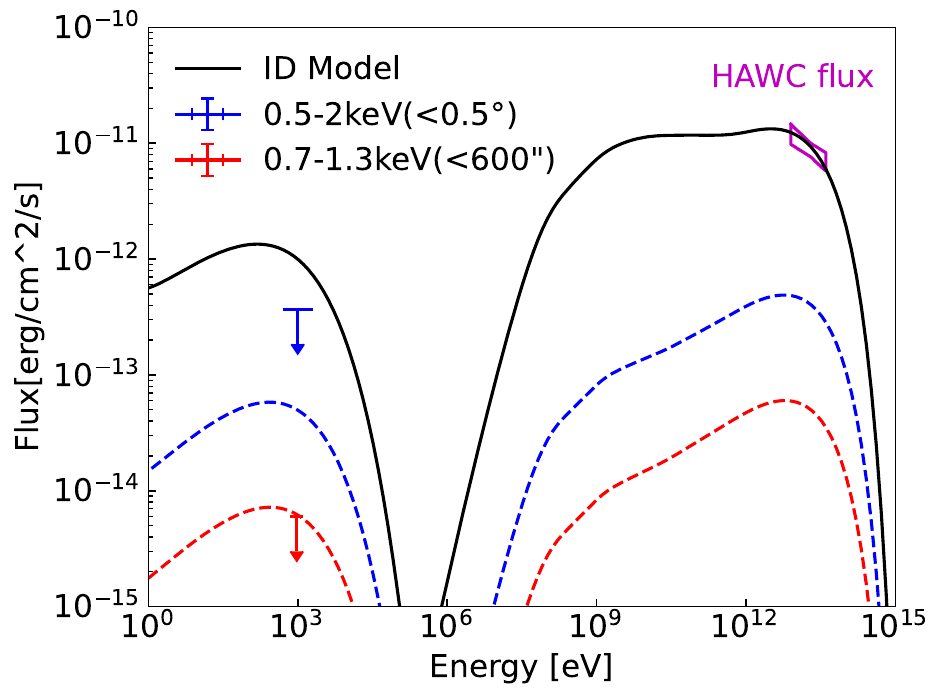}
    \includegraphics[width=0.32\columnwidth]{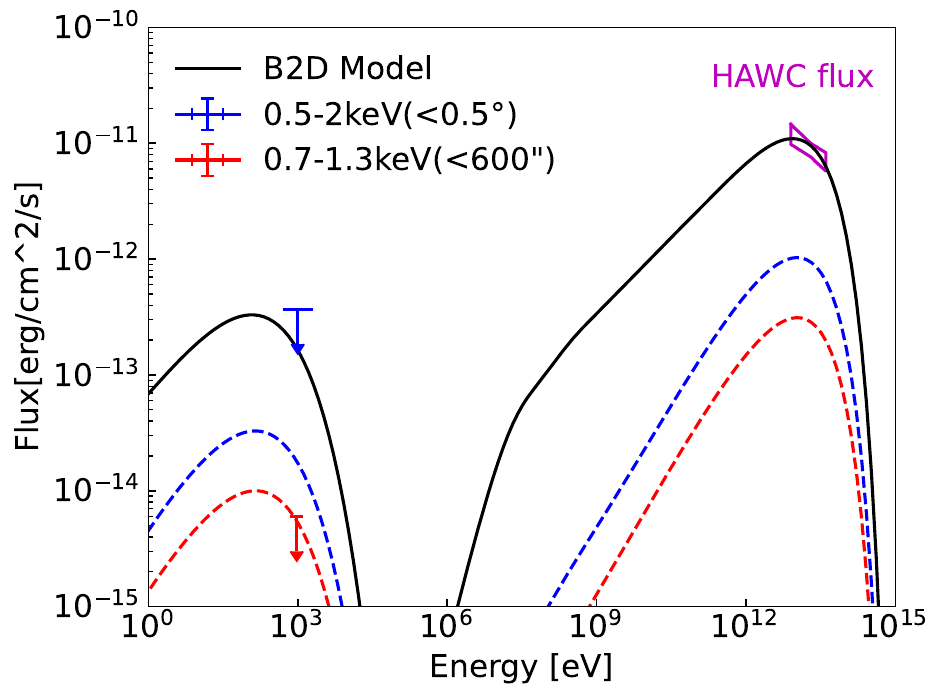}
    \includegraphics[width=0.32\columnwidth]{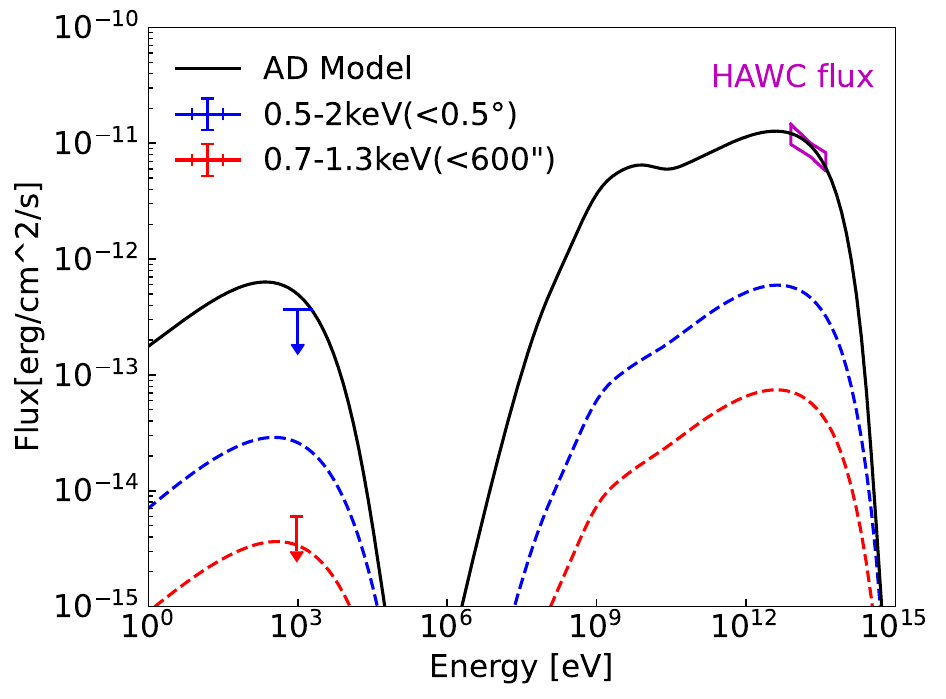}
    }
    \caption{The expected Surface brightness profile in TeV and the Energy spectrum of Geminga with ID Model, B2D Model and AD Model. Note that the red and blue dashed lines correspond to the flux within 600$^{\prime\prime}$ and within $0.5^{\circ}$ region, respectively.}
    \label{fig5}
\end{figure}
\begin{figure}[]
    \centering
    \includegraphics[width=0.96\columnwidth]{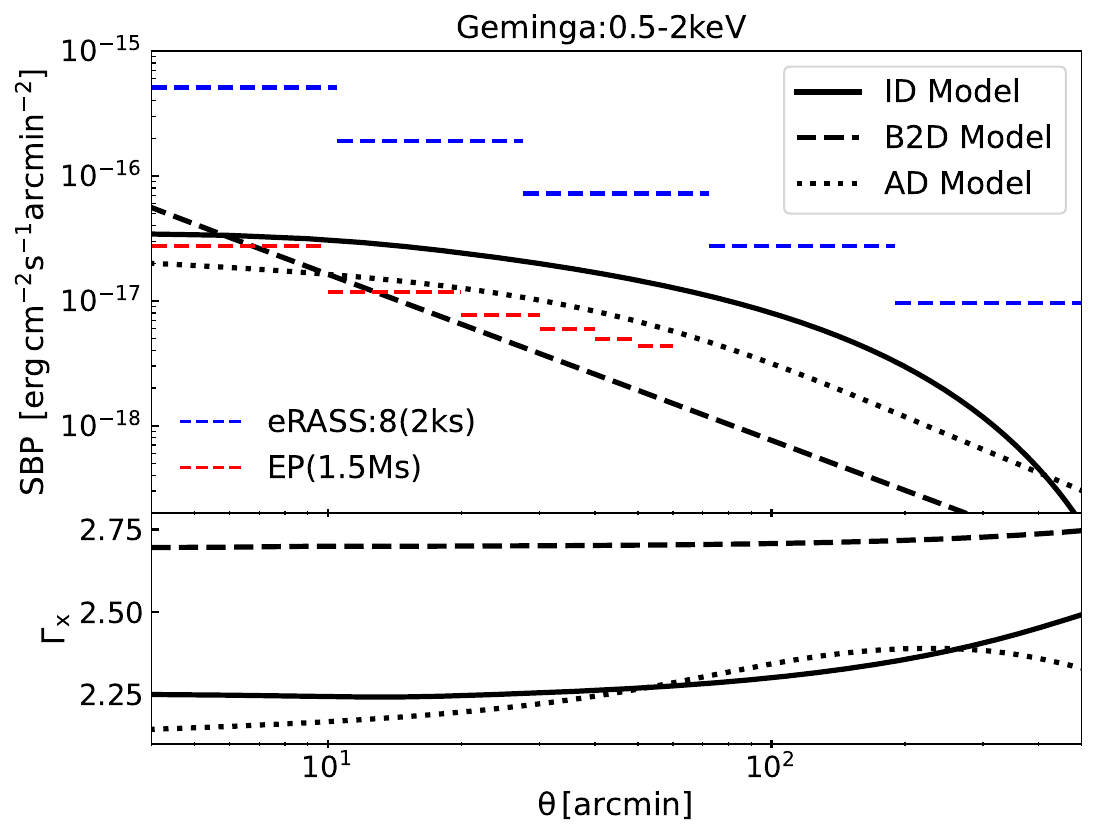}
    \caption{The predicted X-ray intensity and photon index for Geminga with different models. The blue line: the sensitivity of eROSITA all sky survey. The red line: the sensitivity of EP with 1.5\,Ms exposure.}
    \label{fig6}
\end{figure}

\end{document}